\documentclass[manuscript]{aastex}

\shorttitle{Pluto and Eris}
\shortauthors{Tegler et al.}

\begin{document}


\title{Methane and Nitrogen Abundances On Pluto and Eris}


\author{S. C. Tegler\altaffilmark{1,2} and D. M. Cornelison\altaffilmark{3}}
\affil{Department of Physics and Astronomy, Northern Arizona University,
    Flagstaff, AZ, 86011}
\email{Stephen.Tegler@nau.edu, David.Cornelison@nau.edu}

\author{W. M. Grundy\altaffilmark{1}}
\affil{Lowell Observatory,  Flagstaff, AZ, 86001 }
\email{W.Grundy@lowell.edu}

\author{W. Romanishin\altaffilmark{2}}
\affil{Department of Physics and Astronomy, University of Oklahoma, Norman, OK, 73019}
\email{wjr@nhn.ou.edu}

\author{M.R. Abernathy\altaffilmark{4}, M.J. Bovyn, J.A. Burt\altaffilmark{5}, D.E. Evans, C.K. Maleszewski\altaffilmark{1,6},  Z. Thompson\altaffilmark{7}}
\affil{Department of Physics and Astronomy, Northern Arizona University,
    Flagstaff, AZ, 86011}

\author{F. Vilas\altaffilmark{}}
\affil{MMT Observatory, University of Arizona, Tucson, AZ, 85721 }
\email{fvilas@mmto.org}


\altaffiltext{1}{Visiting Astronomer, MMT Observatory. Observations reported here were obtained
at the MMT Observatory, a joint facility of the University of Arizona and the Smithsonian Institution }
\altaffiltext{2}{Visiting Astronomer, Steward Observatory 2.3 m Telescope.}
\altaffiltext{3}{Present Address: Department of Physics, Astronomy, and Materials Science, Missouri State University, Springfield, Missouri, 65897 .}
\altaffiltext{4}{Present Address: Department of Physics and Astronomy, University of Glasgow, Glasgow, G12 8QQ, Scotland.}
\altaffiltext{5}{Present Address: Department of Astronomy and Astrophysics, University of California, Santa Cruz, CA, 95064.}
\altaffiltext{6}{Present Address: Lunar and Planetary Laboratory, University of Arizona, Tucson, AZ, 85721.}
\altaffiltext{7}{Present Address: Department of Physics, Oregon State University, Corvallis, OR, 97331.}


\begin{abstract}
We present spectra of  Eris from the MMT 6.5 meter  telescope and Red Channel Spectrograph (5700$-$9800 \AA; 5 \AA\  pix$^{-1}$) 
on Mt. Hopkins, AZ, and of Pluto from the Steward Observatory 2.3 meter telescope and Boller and Chivens spectrograph (7100$-$9400 \AA; 2 \AA\  pix$^{-1}$) on Kitt Peak, AZ. In addition, we present laboratory transmission spectra of methane-nitrogen and methane-argon ice mixtures.  By anchoring our analysis in methane and nitrogen solubilities in one another as expressed in the phase diagram of  \cite{py83}, and comparing methane bands in our Eris and Pluto spectra and methane bands in our laboratory spectra of methane and nitrogen ice mixtures, we find Eris'  bulk methane and nitrogen abundances are $\sim$ 10\%  and  $\sim$ 90\%  and Pluto's bulk methane and nitrogen abundances are $\sim$ 3\% and $\sim$ 97\%.  Such abundances for Pluto are consistent with values reported  in the literature.  It appears that the bulk volatile composition of Eris is similar to the bulk volatile composition of Pluto. Both objects appear to be dominated by nitrogen ice. Our analysis also suggests, unlike previous work reported in the literature, that the methane and nitrogen stoichiometry is constant with depth into the surface of Eris.  Finally, we point out that our Eris spectrum is also consistent with  a laboratory ice mixture consisting of 40\% methane and 60\% argon. Although we cannot rule out an argon rich surface, it seems more likely that  nitrogen is  the dominant species on Eris because the nitrogen ice 2.15 $\mu$m band is seen in spectra of Pluto and Triton.

\end{abstract}





\section{Introduction}

Pluto and Eris are the largest known icy dwarf planets in the Kuiper belt,  and they both exhibit  methane ice  bands in their optical and near-infrared spectra \citep{owe93,gru96,bro05,lic06}. In addition, Pluto exhibits nitrogen and carbon monoxide ice bands \citep{owe93,dou99}; however, there are no detections of nitrogen and carbon monoxide bands in spectra of Eris. Pluto and Eris have sufficient mass and are sufficiently cold that they are able to retain volatiles in their atmospheres, and hence on their surfaces, over the age of the Solar System \citep{sch07}. 

The detection and analysis of methane, carbon monoxide, and nitrogen ice bands in Pluto's near-infrared spectra has resulted in a number of measurements for Pluto's stoichiometry. \cite{owe93} found a methane 
abundance of 1.5 \%, a carbon monoxide abundance of 0.5\%,  and a nitrogen abundance of 98\%. \cite{dou99} reported a methane abundance of 0.5\%, a carbon monoxide abundance of 0.1\%,  and a nitrogen abundance over 99\%. \cite{olk07} found a methane abundance of 0.36\%. Nitrogen ice 
dominates methane and carbon monoxide ice on Pluto's surface.

Because nitrogen and carbon monoxide bands have yet to be detected in spectra of Eris, it is more difficult to estimate Eris' stoichiometry. One way to qualitatively estimate the methane and nitrogen mixing ratio on Eris is to rely on laboratory studies of methane and nitrogen ices. \cite{qui97} showed that  near-infrared methane absorption bands shift to shorter wavelengths when methane is diluted by nitrogen. We note that \cite{owe93} saw such shifts in their spectra of  Pluto four years before the laboratory work. \cite{bru08} found increasing shifts for methane bands with increasing nitrogen content. Eris' methane bands have smaller shifts than Pluto's methane bands, suggesting that nitrogen is not as abundant on Eris as it is on Pluto \citep{bro05,dum07,mer09}.   

In addition to using methane band shifts as a proxy for the overall methane-nitrogen mixing ratio, it is possible to use the shifts to qualitatively constrain the mixing ratio as a function of depth into the surface of Eris.  The average penetration depth of a photon at a particular wavelength is inversely related to the absorption coefficient at that wavelength, e.g. photons corresponding to larger absorption coefficients are absorbed more, preventing them from penetrating as deeply into the surface. In other words, stronger bands in the spectrum of an icy dwarf planet probe, on average, shallower into the surface than weaker bands. So, it is possible to use blue shifts and albedos at maximum absorption of two or more methane ice bands to look for a trend in the mixing ratio as a function of depth into the surface of Eris.

\cite{lic06} obtained optical spectra of Eris and found evidence of a decreasing nitrogen abundance with depth into its surface. Specifically, they found the weaker 7296 \AA\  band had a shift of 0.1 $\pm$ 3 \AA\   and the stronger 8897 \AA\  band had a shift of 15 $\pm$ 3 \AA. They speculated that as Eris moved toward aphelion during the last 200 years, and as its atmosphere cooled, less volatile methane began to condense out first. As the atmosphere cooled further, it became more and more nitrogen rich, until the much more volatile nitrogen could condense out on top of the methane rich ice.

\cite{mer09} obtained optical and near-infrared spectra of Eris and found further evidence for stratification of methane on Eris. Specifically, they found the 7296 \AA\  and 8897 \AA\  bands were shifted by 9 $\pm$ 4 \AA\  and 12 $\pm$ 4 \AA, respectively.  Furthermore, they found the stronger near-infrared methane bands, which do not probe as deep as the optical bands, had smaller shifts than the optical bands. \cite{mer09} suggested a pure methane surface layer, on top of a nitrogen diluted methane layer, on top of a pure methane layer for Eris.  

We found similar shifts for five methane bands in the spectrum of Eris \citep{abe09}. For example, we found the 7296 \AA\ and 8897 \AA\  bands were blue shifted by 7 $\pm$ 1 \AA\  and 5 $\pm$ 2 \AA, respectively. Abernathy et al.  reported a correlation on the edge of statistical significance between blue shift and albedo at maximum absorption for the five bands. The correlation suggested an increasing nitrogen abundance with increasing depth into the surface, i.e. the opposite trend reported by \cite{lic06} and \cite{mer09}. Abernathy et al. put forth an atmosphere-surface model to explain a way to get an increasing nitrogen abundance with depth. 

All of these stratigraphic investigations make the assumption that different methane bands should have the same shift for the same stoichiometry. However, laboratory experiments show that the smaller (larger) the wavelength (wavenumber) of a band the smaller (larger) the shift \citep{qui97}. It is essential to take the different shifts for different bands into account before discussing changes in composition with depth. 

Here, we report new optical spectra of Pluto and Eris as well as new transmission spectra of laboratory ice mixtures. We use the new data to place quantitative constraints on Eris' stoichometry. In addition, we address the issue of whether the composition of Eris changes with depth.  In sections 2 and 3 below, we describe our astronomical observations and laboratory experiments. In section 4, we describe the models we used to derive abundances on Pluto and Eris. In section 5, we discuss our results.

\section{Observations}
\subsection{Pluto}
We obtained optical spectra of Pluto on the night of 2007 June 21
UT with the Steward Observatory 2.3 meter telescope and
Boller and Chivens Spectrograph. We used a 2.5 $\times$ 240 arc sec entrance slit and a 600 g
mm$^{-1}$ grating that provided wavelength coverage of 7100$-$9400 \AA\  
and a dispersion of 2.0 \AA\  pixel$^{-1}$ in first order.
There were high, thin cirrus clouds and the seeing was $\sim$ 1.5 arc sec  throughout the night.  
Pluto was placed at the center of the slit and the telescope was tracked at Pluto's rate. 
At the start of our observations Pluto had an airmass of about 1.6, it transited the meridian
at an airmass of 1.5, and at the end of our observations it had an airmass of about 1.6.

Since the slit width was 2.5 arc sec and the seeing was $\sim$ 1.5 arc sec, we observed Pluto and Charon blended together. Pluto contributed $\sim$ 90\% to the total signal. 

We used the calibration techniques described by \cite{abe09}. Briefly, we bracketed object spectra with HeNeAr spectra in order to obtain an
accurate wavelength calibration.  Telluric and Fraunhofer lines were
removed from the Pluto spectra by dividing each 600-second Pluto spectrum by the
spectrum of a solar analog star, BS 5996.  The airmass difference
between each Pluto spectrum and its corresponding solar analog spectrum
was $<$ 0.1.

We obtained 12 600-second exposures of Pluto. Since we saw no significant difference between the individual exposures  during the night, we summed them to
yield a single spectrum of Pluto with an exposure time of two hours.
In Figure 1, we present the resulting reflectance spectrum of Pluto.
The spectrum was normalized to 1.0 between 8200 \AA\   and 8400
\AA  and moved upward by 0.2 on Figure 1 to facilitate comparison with the spectrum of Eris (see below).  

We checked our wavelength calibration by measuring the centroids of 
ten well-resolved sky emission lines across
our spectrum and compared them to the VLT high$-$spectral resolution sky
line atlas of \cite{han03}. The average difference and standard deviation of the differences for the ten lines were -0.003 \AA\  and 0.15 \AA.   So, it appears our sky lines, and hence our wavelength measurements, are accurate to $\sim$ 0.2 \AA.

\subsection{Eris}
We obtained optical spectra of Eris on the night of 2008 October 3 
UT with the MMT 6.5 meter telescope, Red Channel Spectrograph, and a red sensitive, deep depletion
CCD. We used a 1 $\times$ 180 arc sec entrance slit and a 270 g
mm$^{-1}$ grating that provided wavelength coverage of 5700$-$9800 \AA\  
and a dispersion of 5.0 \AA\  pixel$^{-1}$ in first order.
There were high, thin cirrus clouds and the seeing was variable, 0.8 $-$ 1.6 arc sec .  
Eris was placed at the center of the slit and the telescope was tracked at Eris' rate. 
At the start of our observations Eris had an airmass of 1.75, it transited the meridian
at an airmass of 1.25, and at the end of our observations it had an airmass of 1.89.

Again, we used the calibration techniques described by \cite{abe09}.  We bracketed object spectra with HeNeAr spectra in order to obtain an
accurate wavelength calibration.  Telluric and Fraunhofer lines were
removed from the Eris spectra by dividing each 600-second Eris spectrum by the
spectrum of a solar analog star, SA 93$-$101.  The airmass difference
between each Eris spectrum and its corresponding solar analog spectrum
was $<$ 0.1.

Since we saw no significant difference between the 24 600-second
exposures of Eris, we summed the individual exposures  to
yield a single spectrum with a four hour exposure time. 
In Figure 1, we present the resulting reflectance spectrum.
The spectrum was normalized to 1.0 between 8200 and 8400
\AA. The  absorption bands in the spectrum are due to methane ice and the reddish slope in the 
"continuum" is consistent with  absorption by a tholin-like material.  

To assess the uncertainty in our wavelength calibration, we measured
the wavelengths of  well-resolved sky emission lines in our
spectrum and compared them to the VLT high$-$spectral resolution sky
line atlas of \cite{han03}. Unfortunately, due to the rather low spectral resolution of our spectrum, only the 6300.3 \AA\  and 9376.0 \AA\  sky lines were well-resolved. Our measured centroids  for these two lines were 0.4 and 0.8 \AA\  different from the centroids of Hanuschik. So, it appears our sky lines, and hence our wavelength measurements, are accurate to $\sim$ 0.8 \AA.

\section{Laboratory Experiments}
\subsection{Setup}
We performed experiments in the ice spectroscopy laboratory at Northern
Arizona University.  This facility includes a series of enclosed ice
sample cells machined by the Lowell Observatory instrument shop out
of aluminum alloy 7075 stock.  Each cell was designed to be
mounted upon the upward-pointing cold tip of a CTI Cryogenics model
1050C two stage closed cycle helium refrigerator driven by a CTI
9600 compressor.  A 15 mm diameter cylindrical cavity in each 
cell holds the cryogenic ice sample for spectroscopic study.
To permit transmission of the spectrometer beam through the samples,
this cylindrical cavity is capped on each end with a sapphire window
sealed in place with indium gaskets.  Different cells have different
spacing between their windows, enabling transmission spectroscopy through
samples of various thicknesses.  For the work presented here, we used
a cell with a 5 mm thickness. Sample materials were introduced and
removed through a stainless steel flexible fill tube which entered the cell
from above.  

Temperatures were measured using a pair of Lake Shore silicon
diode thermometers, mounted on copper rods press-fitted into the cell.
One of these is located just above the sample cavity and the other just
below the sample cavity.  A pair of 50 ohm nickel-chromium heater wires
was wrapped around each cell and cemented into place with Emerson and
Cuming's thermally conductive StyCast epoxy.  As with the diodes, one
heater was wrapped above and one was wrapped below the sample cavity.  
A Cryo-con
model 44 temperature controller was used to monitor and to control the
temperature of the cell, via proportional, integral, and derivative (PID)
closed-loop control of both heaters.  By adjusting the relative amounts of
power going into the upper and lower heaters, it was possible to induce a
small vertical thermal gradient of up to approximately 1 $^{\circ}$K/cm across the
sample (2 $^{\circ}$K between diode thermometers, with more power into the upper heater corresponded to a larger vertical
temperature gradient).  Temperature control stability with the Cryo-con
44 was excellent, with unintended temperature fluctuations generally
below 0.01 degree.  There is a larger absolute uncertainty of less than
1 K  on the sample temperature, owing
to uncertainties in the calibration of the diode thermometers as well
as the fact that they are mounted above and below the sample cavity,
rather than directly sampling the ice.  
We were able to cool samples down to $\sim$ 40$^{\circ}$K.

The cold head and cell were
mounted within an evacuated and turbo-pumped stainless steel cylinder
for insulation, and to eliminate contaminants which could condense on the
exterior of the cell.  The existence of the outer cylinder necessitates
an additional pair of windows to pass the spectrometer beam.  

We used a Perkin Elmer Nexus 670 Fourier transform infrared (FTIR) spectrometer to generate an external beam, and glass lenses to focus the beam within the cell and again onto a silicon detector. All
experiments covered a frequency (wavelength) range from 
$\sim$15000 cm$^{-1}$ (6667 \AA)  to $\sim$ 9000 cm$^{-1}$ (11111 \AA) at a spectral resolution of $\sim$ 0.5 cm$^{-1}$, although lower resolution was used in a few cases, where the signal to noise ratio was poor.
At least 128 scans were taken in each experiment. In some experiments,  512 scans were taken to improve the signal to noise ratio. We present a schematic diagram of our lab set up in Figure 2. 

\subsection{Sample Preparation}
Three separate gas bottles could be connected to a stainless-steel manifold through standard
VCR vacuum fittings. In the experiments reported here, we used methane, nitrogen, and argon 
bottles.  The purities of the gasses were 99.999\% for methane and 99.9\% for nitrogen and argon.
A gas was released into the manifold, and then pumped out by a turbo pump and monitored by a 1000 Torr 
manometer. The purging process was repeated five to ten times. After the purges, the manifold was filled
with a gas mixture dependent on the particular experiment.  The cell was cooled to a temperature slightly above the melting point of the ice. In the case of mixtures, the temperature depended on the mixing ratio of the gases. The gases were then bled into the cell, at which time a liquid could be clearly seen to fill the cell.
When the level of liquid was at least half the total volume of the cell, we stopped the gas flow. The manifold did not have a large ballast volume, and two fills of the manifold were usually required to achieve a satisfactory volume in the cell. After filling the cell with liquid,we slowly ramped the temperature down to below the melting point. During the ramp down, a gradient of 2 $^{\circ}$K was typically kept on the cell.
The target temperature was not always known, as the melting point for binary alloys varies a good deal.
During the ramp down, the ice was periodically observed for freezing. If during the freezing process, the ice
became optically thick, the process was stopped and modified until an optically clear ice could be made.
Typically, we melted the optically thick ice and then reduced the size of the gradient across the sample and ramped the temperature down more slowly than the previous unsuccessful attempt to grow a clear ice. 
It should be noted that making an ice is trivial; however, making an optically clear, two-component ice is a difficult process.

Spectra were taken during sample preparation as follows. A background spectrum was taken before we filled the cell with liquid. Another spectrum was taken after we filled the cell with liquid. After growing a clear ice, we ramped the temperature down further, and spectra were taken at regular temperature intervals.

\subsection{Data Analysis}
Our spectra were analyzed as follows.  Each spectrum was fitted with a LOWESS filter \citep{cle79} to take 
out some minor high frequency noise. 
We assumed that the absorption in our sample was governed by the Beer-Lambert Law, and so
we computed Lambert absorption coefficient spectra, $\alpha(\nu)$. 

We measured the methane and nitrogen stoichiometry of a sample as follows.  We integrated a band in
a sample and the corresponding band in the pure methane sample to obtain integrated absorption coefficients. Next, we divided the integrated absorption coefficient of the sample by the integrated coefficient of pure methane. The result gave us the fraction of methane in our sample.  We repeated the procedure for the other bands in the spectrum.  From the scatter of the values for a sample, we found our solid-phase concentrations have an uncertainty of $\sim$ 5\% of the reported concentration. We found most of the uncertainty came from fitting backgrounds to the spectra. 

We computed the spectral shift for each band using a cross-correlation technique. We shifted a band in the pure methane spectrum relative to the same band in the sample spectrum over a range of shifts. We computed $\chi^2$ for each shift. We fitted a parabola to a plot of $\chi^2$ vs. shift, the minimum of the parabola gave us the shift for the band.  Again, we found most of the uncertainty came from fitting background to the spectra. We found our shifts have an uncertainty of $\sim$ 10\% of the reported shift.

\subsection{Methane-Nitrogen Results}
Next, we present our results for methane-nitrogen ice mixtures. In Figure 3, we plot the 11240 cm$^{-1}$ (8897 \AA)  band in the spectrum of a pure methane ice sample (black line).
In addition, we plot the same band  in the spectrum of a 6 \% methane and a 94\% nitrogen
sample (red line). Both samples were at T $=$ 60 $^{\circ}$K. Our spectrum for pure methane is in
excellent agreement with \cite{gru02}.  From cross correlation experiments, we found
the 11240 cm$^{-1}$ (8897 \AA) band in the spectrum of the 6\% methane sample was  blue shifted by 26 cm$^{-1}$  (21 \AA) relative to the  band of the pure methane sample. In Figure 4, we plot the absorption coefficient spectrum for 
the 13706 cm$^{-1}$ (7296 \AA) band. We found the band in the spectrum of the 6\% sample was blue shifted by 31 cm$^{-1}$ (17 \AA) relative to the band of the pure methane sample.  

In Table 1, we present our blue shifts for samples with methane concentrations of 2, 6, 24, and 99\% 
relative to the bands of pure methane samples. In Figure 5, we plot the values in Table 1. Two trends are apparent in the table and figure. First, the larger the nitrogen concentration, (and so the smaller the methane concentration), the larger the blue shift.  Second, the higher frequency  band always had a larger blue shift than the lower frequency band.  \cite{qui97} found the same
trend of increasing blue shift with increasing frequency in infrared bands. Finally, we note that \cite{qui97} found the 11240 cm$^{-1}$ band had a blue shift of 33 cm$^{-1}$ in spectra of samples with methane abundances of 0.1, 0.25, 0.8, and 2\% at 44 $^{\circ}$K. Their shift is in good agreement with our value of 29 cm$^{-1}$ for a 2\% methane sample (see Table 1).

From Figure 5, it is clear that we did not study intermediate compositions. Methane rich gas mixtures did not easily go into the liquid phase. We had to dramatically increase the gas pressure and substantially lower the temperature, and then almost no liquid would condense, even at methane concentrations above 95 \%.  This effect was seen at all times, and is mostly likely related to deviations from ideal mixtures. When we made a liquid, the ice grown was not optically clear. 

Besides studying the effect of concentration on shift, we wanted to perform an experiment to study the effect
of temperature on shifts. Unfortunately, upon lowering the temperature the ice went through a phase transition (see Section 4.2), and it developed strong scattering characteristics. 

\subsection{Methane-Argon  Results}
Next, we present our results for methane-argon ice mixtures. In Figure 6, we plot the 11240 cm$^{-1}$ (8897 \AA)  band in the spectrum of a pure methane ice sample (same black line as in Figure 3).
In addition, we plot the same band  in the spectrum of a 9\% methane and a 91\% argon sample (red line).
Both samples were at  T $=$ 60 $^{\circ}$K. From cross correlation experiments, we found
the 11240 cm$^{-1}$ (8897 \AA) band in the spectrum of the 9\% methane sample was  blue shifted by 38 cm$^{-1}$ (30 \AA) relative to the  band of the pure methane sample. 

In Table 2, we present our blue shifts for bands in the spectra of samples with methane concentrations of 9, 15, 41, and 96\% methane relative to the corresponding bands of pure methane samples. In Figure 7, we plot the values in Table 2. As in the methane-nitrogen samples, two trends are apparent in the figure and table. First, the larger the argon concentration, (and so the smaller the methane concentration), the larger the blue shift.  Second, the higher frequency band always had a larger blue shift than the lower frequency band . 

Besides studying the effect of concentration on shifts, we performed an experiment to study the effect of 
temperature on shifts despite the ice developing strong scattering characteristics. Specifically, we cooled the 9\% methane 91\% argon ice from $\sim$ 60 $^{\circ}$K down to about 40 $^{\circ}$K. In Figure 6, we plot the 11240 cm$^{-1}$ (8897 \AA) band at T $\sim$ 40 $^{\circ}$K (blue line). It is clear that the single peak is split into two peaks. We warmed the sample up to T $\sim$ 60 $^{\circ}$K, and once again obtained a single peak (red line in Figure 6).  We discuss
the reason for the change in band profile with temperature in section 4.2.

\section{Models}
Next, we describe the models we used to derive methane and nitrogen abundances on Pluto and Eris. First, we describe a pure methane model. Then, we describe a binary mixture model. 

\subsection{Pure Methane}
In order to establish continuity with previous work, 
we  compared our spectra of Pluto and Eris to
spectra of pure methane ice at 30$^{\circ}$K .
We used Hapke theory to transform laboratory optical
constants of pure methane ice at 30$^{\circ}$K \citep{gru02} into
spectra suitable for comparison to the Pluto and Eris spectra. In particular, Hapke
theory accounts for
multiple scattering of light within a surface composed of particulate
ice \citep{hap93}.  We used Hapke model parameters of h $=$
0.1, B$_o$ $=$ 0.8, $\overline{\theta}$ $=$ 30$^{\circ}$, P(g) $=$ a
two component Henyey-Greenstein function with 80$\%$ in the forward
scattering lobe and 20$\%$ in the back scattering lobe, and both lobes
had asymmetry parameter a $=$ 0.63. 
Our implementation of the Hapke model used two particle sizes and a tholin-like
absorber.  We explored parameters specific to our implementation, i.e. particle diameters (D$_1$ and D$_2$), percent of large particles D$_1$ by volume (f), and tholin-like absorption (P$_1$ and P$_2$), with Monte Carlo techniques (as described in more detail by Abernathy et al. 2009).
For example, we performed cross-correlation
experiments between 10,000 Hapke model spectra and the portion of Pluto's spectrum
containing the 8897 \AA\  methane band. 
We shifted each Hapke model spectrum in 2 \AA\   steps
between $-$30 \AA\   and $+$0 \AA, calculated an adjusted $\chi^2$ goodness
of fit, R, for each shift, plotted R vs. shift, and
fitted a parabola to the plot to determine the shift
corresponding to the minimum in R for each model. 

We found a dozen or so of 10,000 Hapke models gave equally good
fits to Pluto's 8897 \AA\  band. The relatively small differences in the
parameters of these models had little
effect on the  blue shift. Our best fit model for the Pluto 8897 \AA\  band had grain sizes 
of D$_1$$=$1.8 cm (89$\%$ by volume) and D$_2$$=$0.1 mm (11$\%$ by volume).
It is important to recognize that our Hapke parameters do not represent a unique fit to Pluto's band;
however, they are plausible values for transparent, pure methane ice
grains and are comparable to Hapke parameters in previous fits of Pluto spectra
\citep{gru01}. 

Our best fit model indicated  Pluto's
8897 \AA\  band is blue shifted by 17 $\pm$ 2  \AA\  relative to pure methane ice.  On 2007 June 21 UT, we were observing the anti-Charon facing hemisphere of Pluto.  \cite{gru96} reported a correlation between blue shift and longitude for Pluto's 8897 \AA\  band, and found a  blue shift of $\sim$ 12 \AA\  for the anti-Charon facing hemisphere. Grundy and Fink measured their shifts in a manner different than we described above and so differences between the two values may not be significant. 

In Figure 8, we present a comparison of Pluto's 8897 \AA\  band (black line) and our (shifted) best fit pure
methane ice Hapke model (red line).  It is clear that Pluto's 8897 \AA\  band is wider than the pure methane 8897 \AA\   band.  We found $\chi^2$ $=$ 14. Unfortunately, noise in our Pluto spectrum prevented us from analyzing any other methane bands. 

We calculated the uncertainty in our blue shift as follows.  
The best fit (shifted) model was subtracted from
Pluto's 8897 \AA\  band, giving us the noise in the astronomical
spectrum. We note that the noise in the astronomical spectrum
dominates noise in the Hapke model spectrum. Next, we calculated the
standard deviation of the noise. Then, we applied Gaussian noise with
the same standard deviation to the model spectrum. Next, we applied
our cross correlation technique to the noisy model and the original
model spectrum. We generated 1000 different noisy model spectra and
repeated the cross correlation experiment for each noisy model and the
original model spectrum.  A histogram of the resulting 1000 shifts was
fit with a Gaussian distribution, the standard deviation of the fit
giving the uncertainty in our shift measurement, 2.0 \AA.

We repeated the analysis described above on our Eris spectrum. In particular, we fit pure methane Hapke models to the 7296 \AA, 8691 \AA, and the 8897 \AA\  methane ice bands. In Table 3, we present our blue shift measurements for these bands. We point out that the Eris shifts in Table 3 are in good agreement with our previously reported shifts for these bands \citep{abe09}. In Figures 9 and 10, we present our Eris 8897 \AA\  and 7296 \AA\ bands (black line) and our (shifted) best fit pure methane Hapke model (red line).  We found $\chi^2$ $=$ 10 for the 8897 \AA\  band and $\chi^2$ $=$ 5 for the 7296 \AA\  band.

\subsection{Binary Mixture}
We began thinking about the mechanism responsible for the creation of the double peak in the 11240
cm$^{-1}$ (8897 \AA) band upon cooling the methane-argon sample down (Figure 6), and
it led us to a new model. Examining the methane-argon phase diagram of \cite{gmb69} that we display in Figure 11, we found that cooling the sample down resulted in changing the structure of the ice from a soluble solution with methane and argon molecules mixed throughout the ice to a solution of limited solubility where we had highly methane rich regions and highly argon rich regions.  The two peaks in the band came from the two compositionally distinct regions. Greer, Myer, \& Barrett  reported that at T $\sim$ 40 $^{\circ}$K, the solubility limit of methane in argon is $\sim$ 13 \% and the solubility limit of argon in methane is $\sim$ 12 \%.  We believe this is the first time anyone has spectroscopically observed the  change in a methane-argon ice from a face-centered cubic (FCC) solid solution to regions dominated by FCC argon and FCC methane. 

The methane-nitrogen phase diagram of \citep{py83} that we display in Figure 12 is similar to the methane-argon phase diagram.
At T $\sim$ 60 $^{\circ}$K our samples were solid solutions.  At T $\sim$ 40 $^{\circ}$K, methane and nitrogen have limited solubility.  The solubility limit of methane in nitrogen is $\sim$ 2\% and the solubility limit of nitrogen in methane is $\sim$ 4\%. Because of the experimental difficulties of growing an optically clear methane-nitrogen ice and cooling it, we were unable to spectroscopically see the structural  change in a methane-nitrogen ice that we saw in the methane-argon ice. However, the methane-nitrogen phase diagram says it is there.  

At the temperatures of Pluto (T $\sim$ 40$^{\circ}$K) and Eris (T$\sim$30$^{\circ}$K), we expect methane and nitrogen to have limited solubility in one another.  So, we developed an ice model for Pluto and Eris consisting of two components $-$ methane-rich crystals near the nitrogen solubility limit and nitrogen-rich crystals near the methane solubility limit. We point out that \cite{dou99} made use of methane-rich regions and nitrogen-rich regions in modeling their near-infrared spectrum of Pluto. Although we could have added additional parameters to enable the two materials to have different grain sizes, we didn't feel the current data supported additional free parameters, and so we assumed the same grain sizes for both types of materials.  In our model f is the overall methane abundance, $\eta$ is the fraction of methane rich crystals, S$_{N_2}$ is the solubility limit of nitrogen in methane, and S$_{CH_4}$  is the solubility limit of methane in nitrogen. The relationship between these quantities is given by

\begin{eqnarray}
f &=& \eta(1-S_{N_2}) + (1-\eta)S_{CH_4}.
\end{eqnarray}

\noindent Solving for $\eta$, we get

\begin{eqnarray}
\eta & =&{{f-S_{CH_4}}\over{1-S_{CH_4}-S_{N_2}}} .
\end{eqnarray}

\noindent  We combined absorption coefficients of the methane rich crystals ($\alpha_u$), and absorption coefficients of the highly diluted methane crystals  ($\alpha_s$), 

\begin{eqnarray}
\alpha(\nu) & =& \eta(1-S_{N_2})\alpha_u(\nu) + (1-\eta)S_{CH_4}\alpha_s(\nu).
\end{eqnarray}

\noindent We approximated $\alpha_u$ with  pure methane  coefficients, and $\alpha_s$ with  pure methane coefficients blue shifted by amounts observed in highly diluted samples, e.g. 33 cm$^{-1}$  for the 11240 cm$^{-1}$ (8897 \AA) band \citep{qui97}. 

We fitted the Pluto and Eris bands much the same way we fitted them with pure methane models in the above section.  The primary difference between the two models is that the binary mixture model had an additional free parameter, the methane abundance, f.  In Figure 13, we plot our best fit binary mixture model (red line) to Pluto's 8897 \AA\  band (black line). Notice that the binary mixture model provides a  better fit to Pluto's wide 8897 \AA\ band  than the pure methane model. We found $\chi^2$ $=$ 10 for the binary mixture model and $\chi^2$ $=$ 14 for the pure methane model.  In addition, there was no apparent wavelength mismatch between the model and the data. The best fit model had a methane abundance of 3.2 $\pm$ 0.2 $\%$, and hence a nitrogen abundance of 96.8$\%$. 

In Figure 14, we plot our best fit binary mixture model to Eris' 8897 \AA\  band.  Again, there was no apparent wavelength mismatch between the model and the data. The best fit model for Eris  had a methane abundance of 
10$\pm$ 2 $\%$, and hence a nitrogen abundance of 90$\%$.   We found $\chi^2$ $=$ 10 for the binary mixture model and $\chi^2$ $=$ 10 for the pure methane model.
In Figure 15, we plot our best fit binary mixture model to the 7296 \AA\  band. We found $\chi^2$ $=$ 3 for the binary mixture model and $\chi^2$ $=$ 5 for the pure methane model.There was no wavelength mismatch between the model and the data. We found a methane abundance of 8 $\pm$ 2 \%. 
In Table 4, we present the methane abundances corresponding to the best fit models for all three Eris bands. 

We analyzed our Eris bands with a methane-argon ice mixture. We found models with 40 $\pm$ 5 \% methane, and hence 60 \% argon, gave the best fits to Eris'  7296 \AA, 8961 \AA, and 8897 \AA\  bands. 

\section{Discussion}
\subsection{Eris' Stoichiometry}
Previous work suggested that the majority of methane ice on Eris is in a pure form rather than in a highly diluted form with nitrogen \citep{bro05,dum07,mer09} . Such thinking rests on two kinds of observations. First, the nitrogen 2.15$\mu$m  band has yet to be detected in spectra of Eris, but it is seen in spectra of Pluto.   Second, methane band shifts on Eris are smaller than on Pluto. 

Our analysis suggests nitrogen is almost an order of magnitude more abundant than methane on the surface of Eris.  By fitting three methane bands in our Eris spectrum with the binary mixture model, we find a methane abundance of $\sim$ 10 $\%$, and hence a nitrogen abundance of $\sim$ 90 $\%$.   Our model is able to fit all three bands. The lack of a 2.15$\mu$m nitrogen band detection in spectra of Eris does not concern us too much. With an albedo of 70$\%$ \citep{stan08} and a heliocentric distance of 97 AU, we expect a subsolar surface temperature of $\sim$ 30$^{\circ}$K for Eris. Such a temperature is below the T$=$ 35.6 $^{\circ}$K transition temperature between the $\alpha$ and $\beta$ phase of nitrogen. So it is possible nitrogen is in the $\alpha$ phase on Eris rather than the $\beta$ phase as it is on Pluto.  The 2.15 $\mu$ absorption feature in the  $\alpha$ phase is much narrower than for the $\beta$ phase \citep{gru93}. If nitrogen is in the $\alpha$ phase,  it will be more difficult to detect on Eris than on Pluto and it could complicate our analysis.  Specifically, the nitrogen in our experiments was in the $\beta$ phase and  \cite{qui97} found the $\nu_3+\nu_4$ methane band had shifts and profiles that strongly depended on the nitrogen phase whereas the $\nu_1+\nu_4$ methane band had shifts and profiles that were weakly dependent on the nitrogen phase. Our lab setup was not capable of going cold enough to grow nitrogen in the alpha phase and see whether any of our methane bands exhibit a strong nitrogen phase dependence.  Another possibility is that some of the nitrogen ice may be hidden by the presence of scattering inclusions such as voids or wind blown  dust particles of another ice. \cite{gru10} suggested such hidden nitrogen on Triton. Our analysis suggests that the volatile composition of Eris and Pluto may not be so different. 

Our spectra also are consistent with Eris having a surface composed of 60\% Ar and 40\% CH$_4$. It not unreasonable to expect Ar on the surface of Eris, although Ar is 30 times less abundant than N in the Solar System \citep{and89}.  If all of the N were in the form of N$_2$, then Ar would be 15 times less
abundant than N$_2$. If some of the N were locked up in the interior of Eris in the form of NH$_3$,  the Ar abundance at the surface could be further boosted relative to N$_2$. Furthermore, Ar is not destroyed by photolysis and radiolysis as is CH$_4$. Eris could have started with a CH$_4$ dominated surface, but photolysis and radiolysis destroyed much of it, leaving Eris' surface dominated by Ar.  Finally, the atomic mass of Ar is larger than the molecular masses of CH$_4$ and N$_2$, and so Eris would be able to retain Ar better than CH$_4$ and N$_2$ in its atmosphere and hence on its surface. 

\subsection{Eris Stratigraphy}
Previous work looked for a correlation between shift (a proxy for the nitrogen abundance) and albedo of maximum absorption (a proxy for depth into the surface) of methane bands. We and others reported such correlations, the signature of stratification.  \cite{lic06} found  essentially no shift for the weaker 7296  \AA\  band, and a  shift of 15 \AA\  for the stronger 8897 \AA\ band. Such a trend suggested a decrease in the nitrogen abundance with depth into the surface. On the other hand, we reported an increase in nitrogen abundance with depth \citep{abe09}.  In the work we report here, we fit models with the same methane and nitrogen stoichiometry for three bands. Furthermore, if there were a change in the stoichiometry with depth, the weaker bands which probe deeper into the surface and therefore probe a range of mixing ratios, should be broader than the stronger bands which don't probe as deep into the surface. We find no evidence that the weaker 7296 \AA\   and  8961 \AA\  bands are  systematically broader than the 8897 \AA\  band.   In short, we see no evidence of stratification in the ice. 

\subsection{Pluto's Stoichiometry}
There are a range of reported values for the methane and nitrogen stoichiometry of Pluto in the literature. \cite{owe93} reported a methane abundance of 1.5$\%$ and an uncertainty of a factor of two. So the Owen et al. methane abundance could be as large as a 3 $\%$.  \cite{dou99} found a methane abundance of 0.5$\%$, and \cite{olk07} found a methane abundance of  0.36$\%$.

Our methane abundance of 3.2 $\pm$ 0.2 $\%$ overlaps with the values reported by \cite{owe93}. However, our methane abundance is inconsistent with abundances smaller than $\sim$ 2$\%$. At such low abundances, the ice would be dominated by highly diluted methane crystals. So, we should see an 8897 \AA\  band that is as narrow as the pure methane 8897 \AA\ band, but shifted 26 \AA\  (33 cm$^{-1}$) relative to the pure methane band. Clearly, the Pluto band in Figure 8 is much wider than the pure methane band and not shifted enough for highly diluted methane. We point out that the \cite{olk07} methane abundance of 0.36\% is based on 3 $\mu$m data, and so does not probe as deep into the surface as our optical data. Perhaps a change in composition with depth is responsible for the different values.  

\subsection{Future Work}
We hope to obtain near- and mid-infrared spectra of Eris at higher spectral resolutions than reported in the literature (R $>$ 15,000). Such observations will test the stoichiometry and  the lack of stratification in Eris' ice described here. In addition, we hope to obtain additional spectra of Pluto and map its methane and nitrogen abundances as a function of longitude. Such measurements may prove useful in the interpretation of New Horizons observations of Pluto. 

\acknowledgments
We are grateful to the NASA Planetary Astronomy Program for a grant to Northern Arizona University (NNX10AB24G) and the University of Oklahoma (NNX10).  We are grateful to the NASA Planetary Geology and Geophysics Program for a grant to Lowell Observatory (NNX10AG43G), and the Mt. Cuba Astronomical Foundation for a grant to Lowell Observatory.  We thank the NSF REU Program and the NASA Spacegrant Program at
Northern Arizona University for supporting student work on this project. We thank the Director and Telescope Allocation Committee of Steward Observatory for the consistent allocation of telescope time.


\clearpage

\begin{figure}
\epsscale{.80}
\plotone{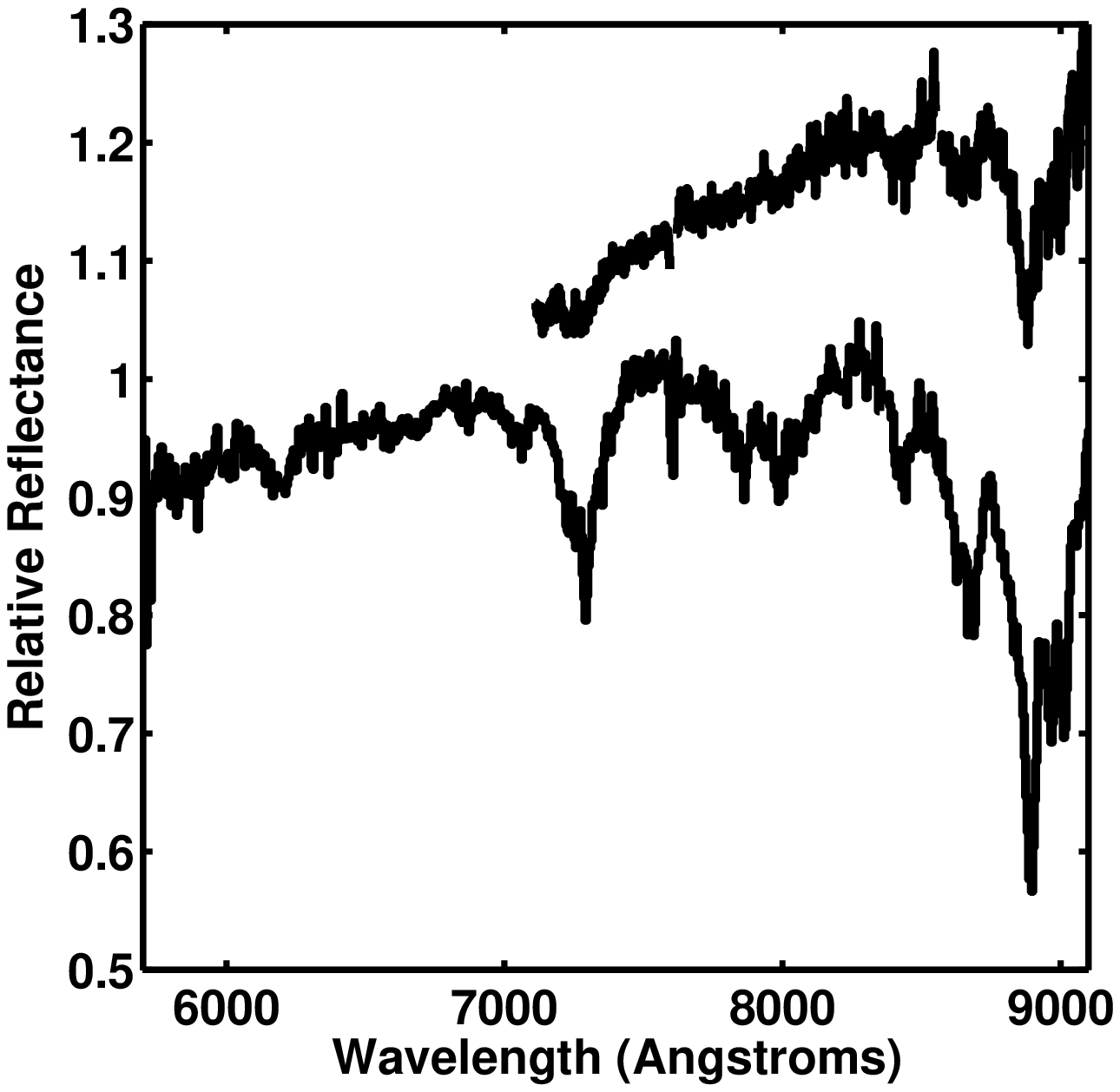}
\caption{Spectrum of Pluto (top) taken on 2007 June 21 UT with the Steward Observatory 2.3 m telescope, normalized to 1 between 8200 \AA\  and 8400 \AA\, and shifted upward
by 0.2 to facilitate a comparison with the spectrum of Eris. Spectrum of Eris (bottom) taken on 2008 October 3 UT with the 6.5 m MMT, and normalized to 1 between 8200 \AA\  and 8400 \AA. The absorption bands in both spectra are due to methane ice and the reddish slope in the "continuum" is consistent with absorption by a tholin-like material.  Eris has deeper methane bands than Pluto.}
\end{figure}

\clearpage

\begin{figure}
\epsscale{.80}
\plotone{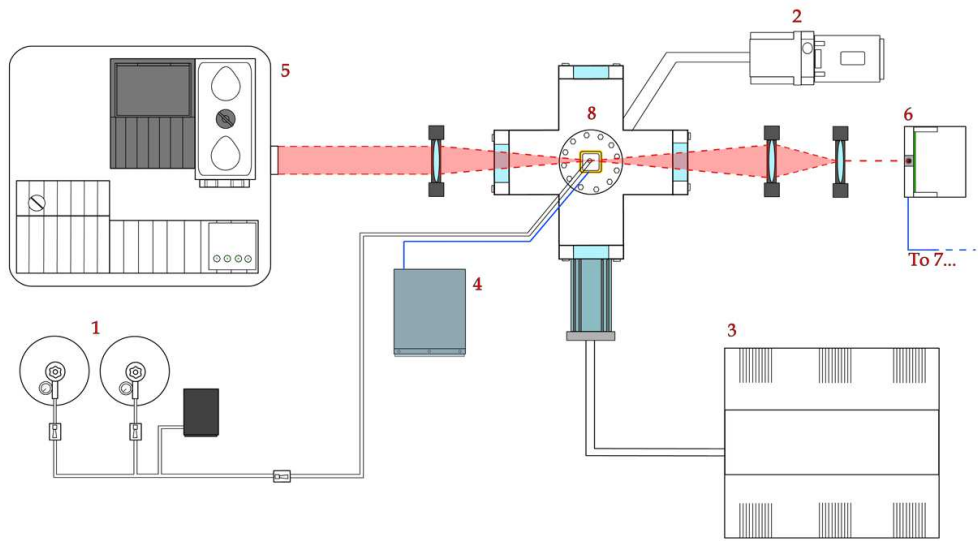}
\caption{A schematic diagram of the set up in our ice laboratory. 1 gas bottles; 2 turbo pump for vacuum that surrounds cell; 3 closed cycle helium refrigerator and compressor; 4 temperature controller and monitor for cell; 5 Fourier transform spectrometer; 6 silicon detector; 7 to data acquisition computer; 8 cell. }
\end{figure}

\clearpage

\begin{figure}
\epsscale{.80}
\plotone{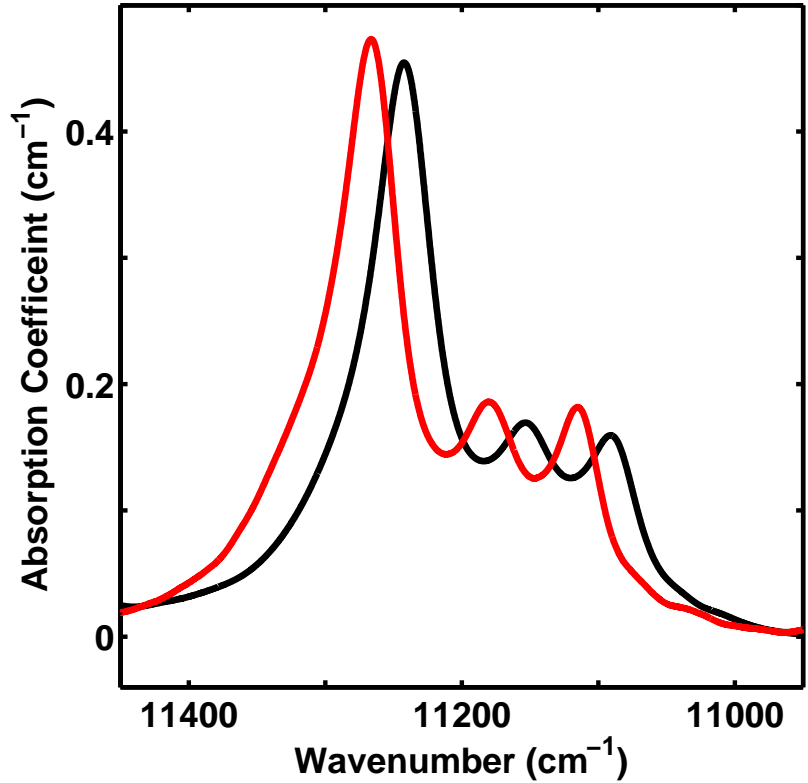}
\caption{The 11240 cm$^{-1}$ (8897 \AA) band in the spectrum of a pure methane ice sample (black line). The same band in the spectrum of a sample of 6\% methane and 94\% nitrogen (red line).  The band corresponding to the highly diluted methane sample (red line) has the same shape as the band corresponding to the pure sample (black line), but it is blue shifted by 26 cm$^{-1}$ (21 \AA). }
\end{figure}

\begin{figure}
\epsscale{.80}
\plotone{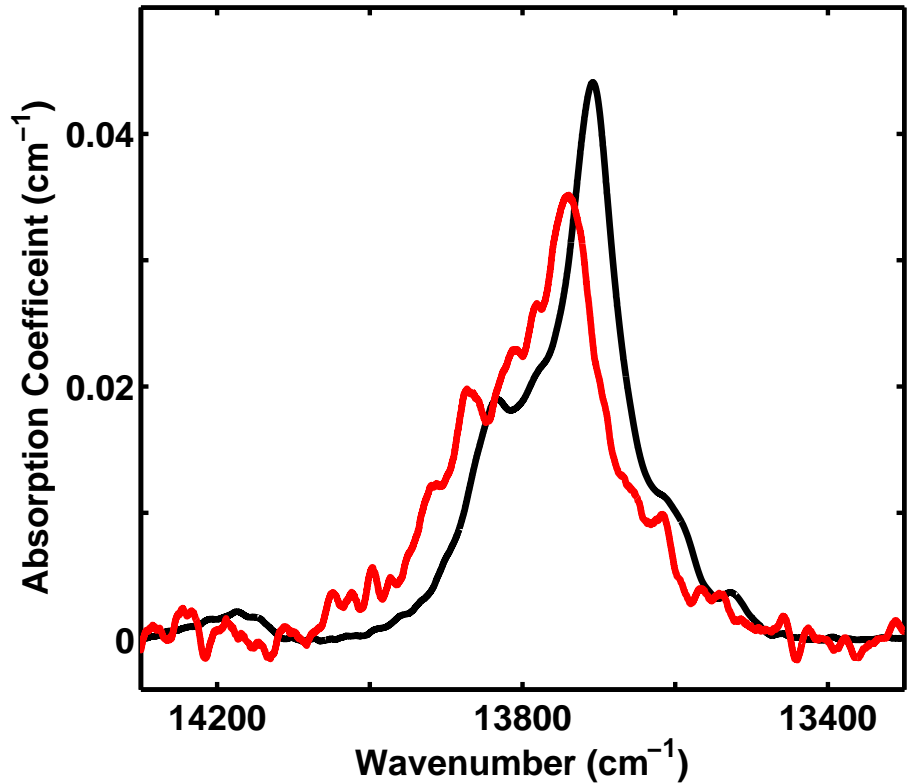}
\caption{The 13706 cm$^{-1}$ (7296 \AA) band in the spectrum of a pure methane ice sample (black line). The same band in the spectrum of a sample of 6\% methane and 94\% nitrogen (red line).  The band corresponding to the highly diluted methane sample (red line) has the same shape as the band corresponding to the pure sample (black line), but it is blue shifted by 31 cm$^{-1}$ (17 \AA). The low signal to noise ratio of the diluted band is due to the weak absorption efficiency of the band and the low methane abundance of the sample.}
\end{figure}

\begin{figure}
\epsscale{.80}
\plotone{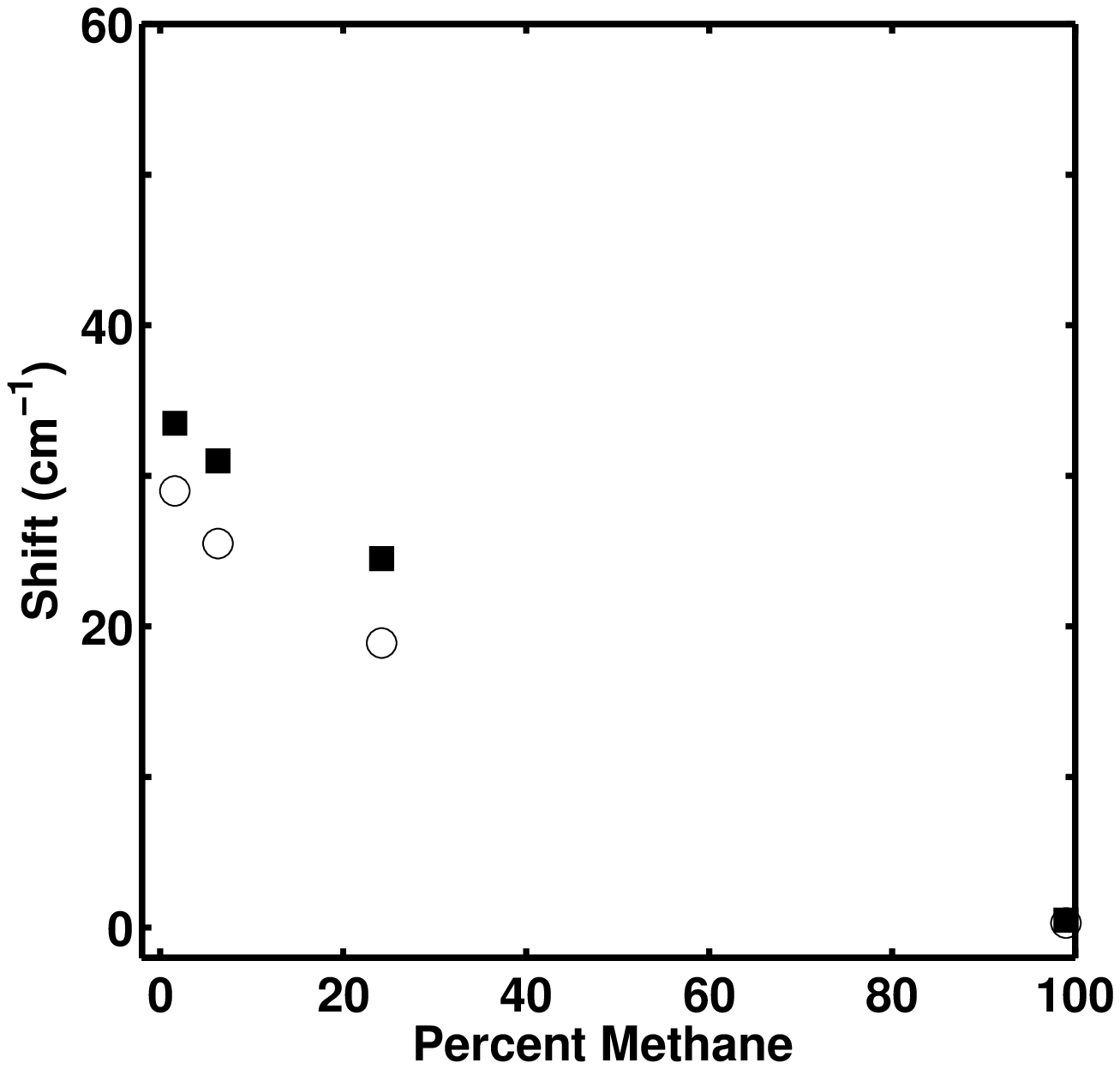}
\caption{Blue shift of the 11240 cm$^{-1}$ (8897 \AA) band (open circles) and the 13706 cm$^{-1}$ (7296 \AA ) band (filled squares) vs. methane abundance. The larger the nitrogen abundance (smaller the
methane abundance), the larger the blue shift of the band.  For the same methane concentration, the 13706 cm$^{-1}$ band has a larger shift than the 11240 cm$^{-1}$ band.}
\end{figure}

\begin{figure}
\epsscale{.80}
\plotone{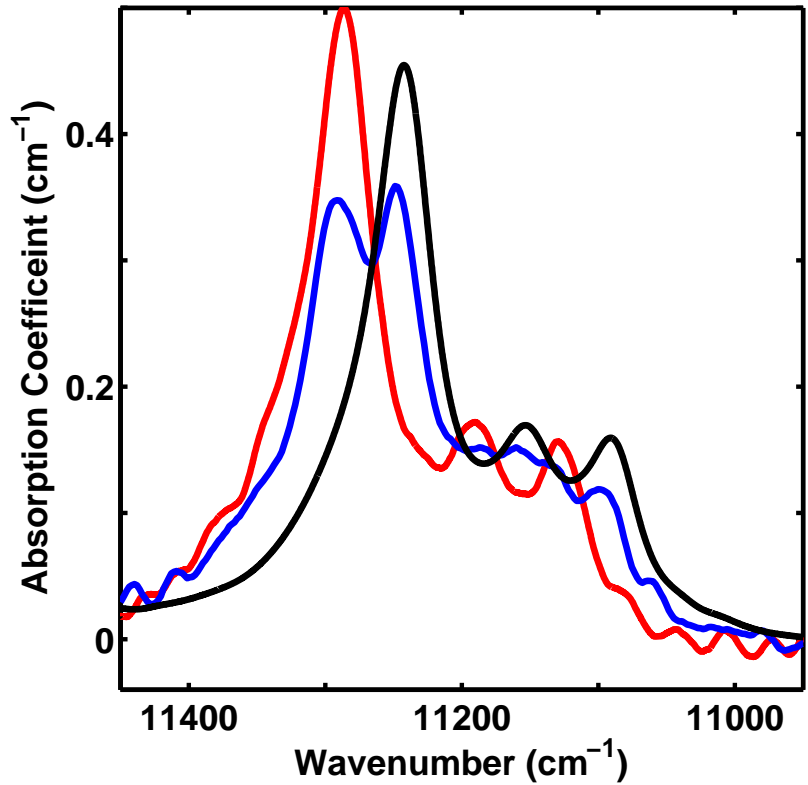}
\caption{The 11240 cm$^{-1}$ (8897 \AA) band in the spectrum of a pure methane ice sample at T $\sim$ 60 $^{\circ}$K (black line). The same band in the spectrum of a sample of 9\% methane and 91\% argon at T $\sim$ 60 $^{\circ}$K (red line).  The highly diluted methane band, but at T $\sim$ 40 $^{\circ}$K (blue line). At the higher temperature, the methane and argon molecules are soluble and methane and argon molecules are mixed throughout the ice. At the lower temperature, the methane and argon are not soluble and there are regions of nearly pure methane and regions of highly dilute methane giving rise to the two peaks in the band.}
\end{figure}

\begin{figure}
\epsscale{.80}
\plotone{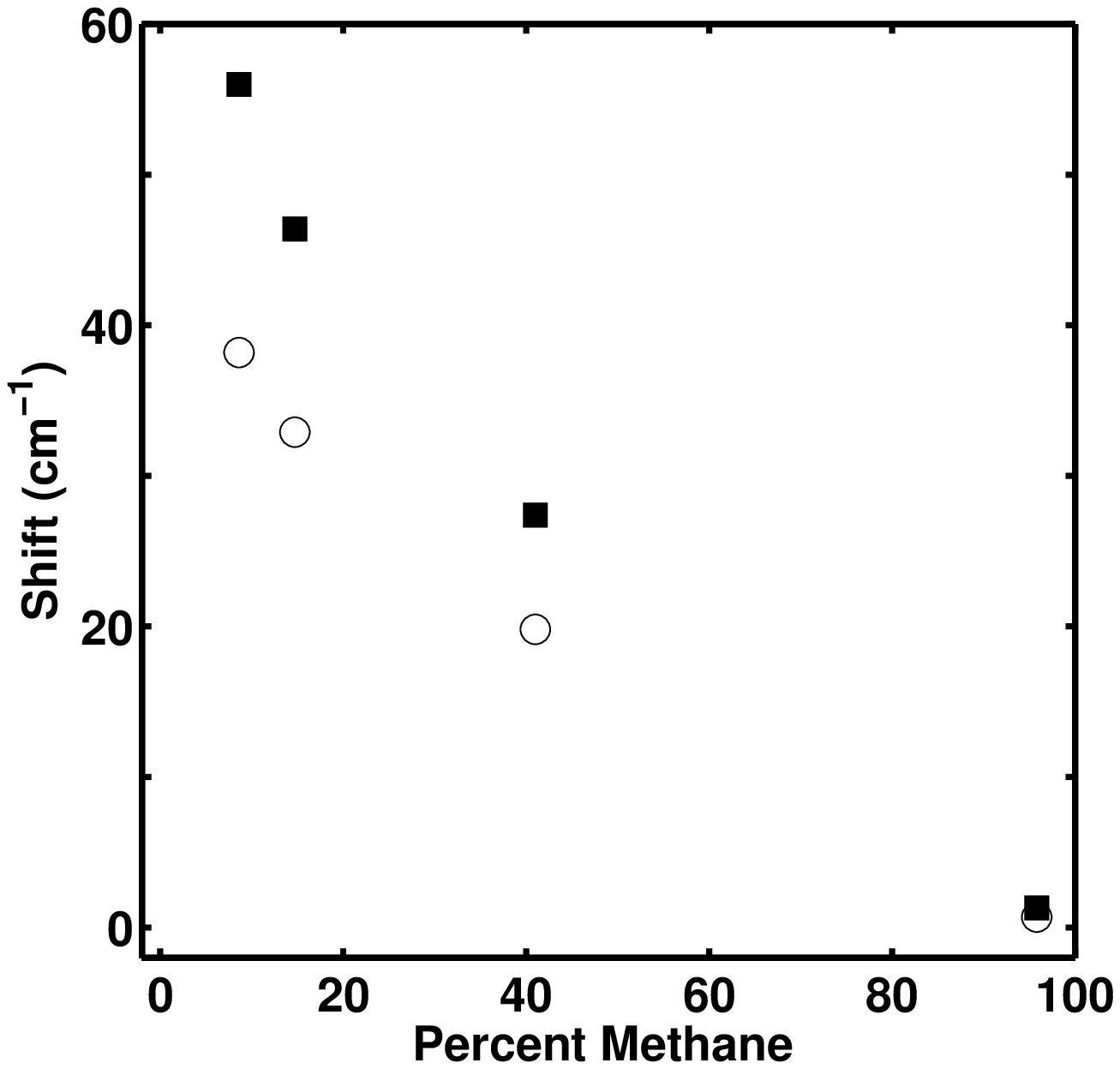}
\caption{Blue shift of the 11240 cm$^{-1}$ (8897 \AA) band (open circles) and the 13706 cm$^{-1}$ (7296 \AA ) band (filled squares) vs. methane abundance. The larger the argon abundance (smaller the
methane abundance), the larger the blue shift of the band.  For the same methane concentration, the 13706 cm$^{-1}$ band has a larger shift than the 11240 cm$^{-1}$ band.}
\end{figure}

\begin{figure}
\epsscale{.80}
\plotone{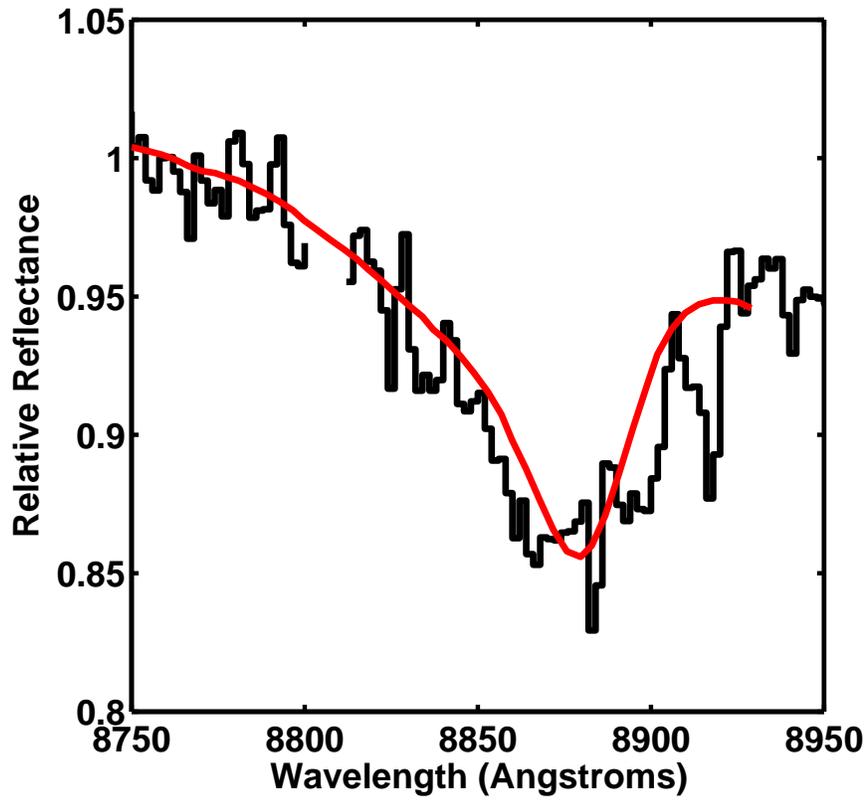}
\caption{Pluto's 8897 \AA\  methane band taken on 2007 June 21 UT with the Steward Observatory 2.3 m telescope (black line), and the best fit pure methane Hapke model (red line). Blue shifting the pure methane band by 17 $\pm$ 2 \AA\  gave the best fit to Pluto's band. Notice that the Pluto band has a wider profile than the pure methane band. $\chi^2$ $=$ 14.}
\end{figure}

\begin{figure}
\epsscale{.80}
\plotone{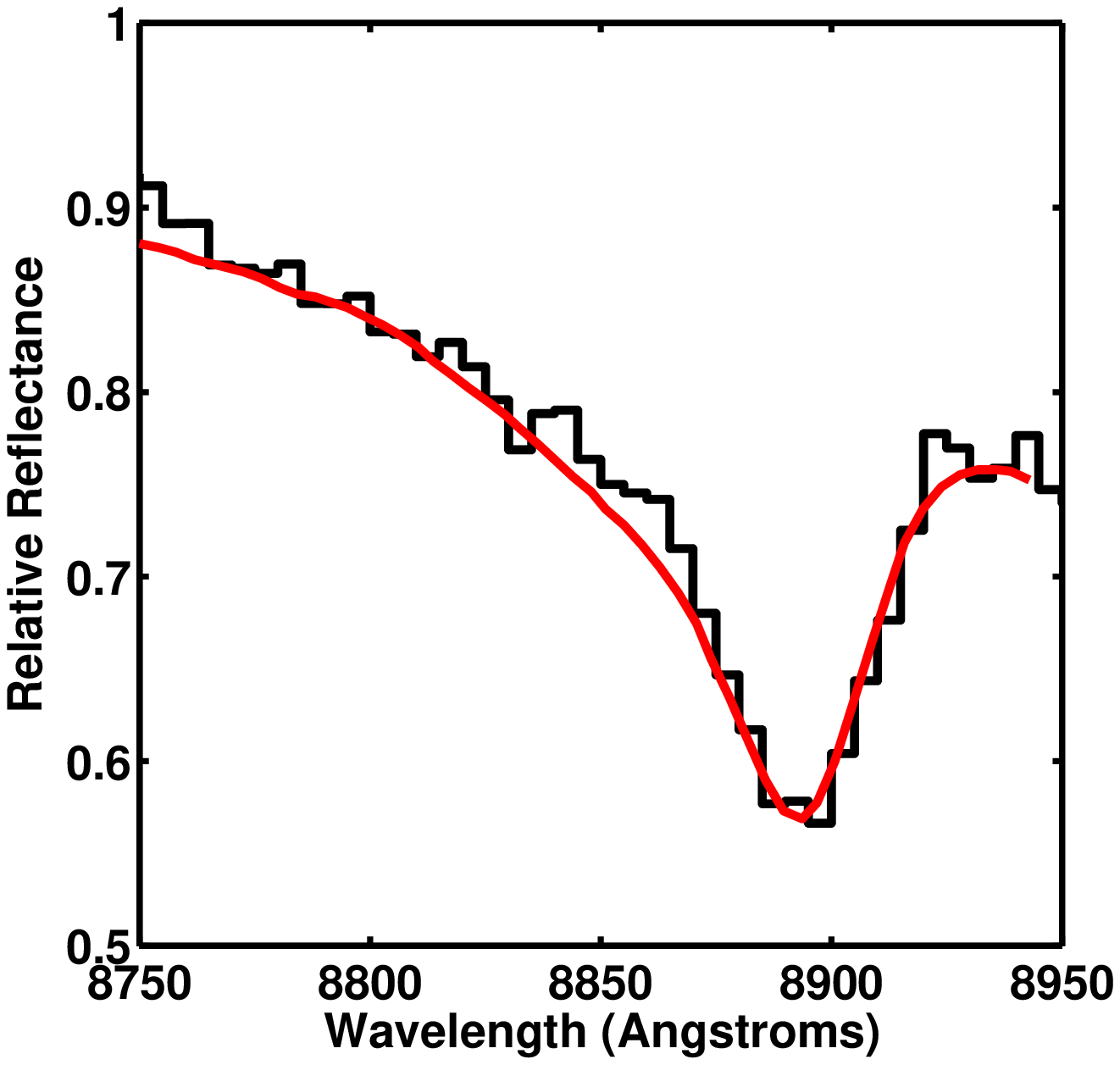}
\caption{Eris' 8897 \AA\  methane band taken on 2008 October 3 UT with the MMT 6.5 m telescope (black line), and the best fit pure methane Hapke model (red line). Blue shifting the pure methane band by 4.0 $\pm$ 0.6 \AA\  gave the best fit to Eris' band. $\chi^2$ $=$ 10}
\end{figure}

\begin{figure}
\epsscale{.80}
\plotone{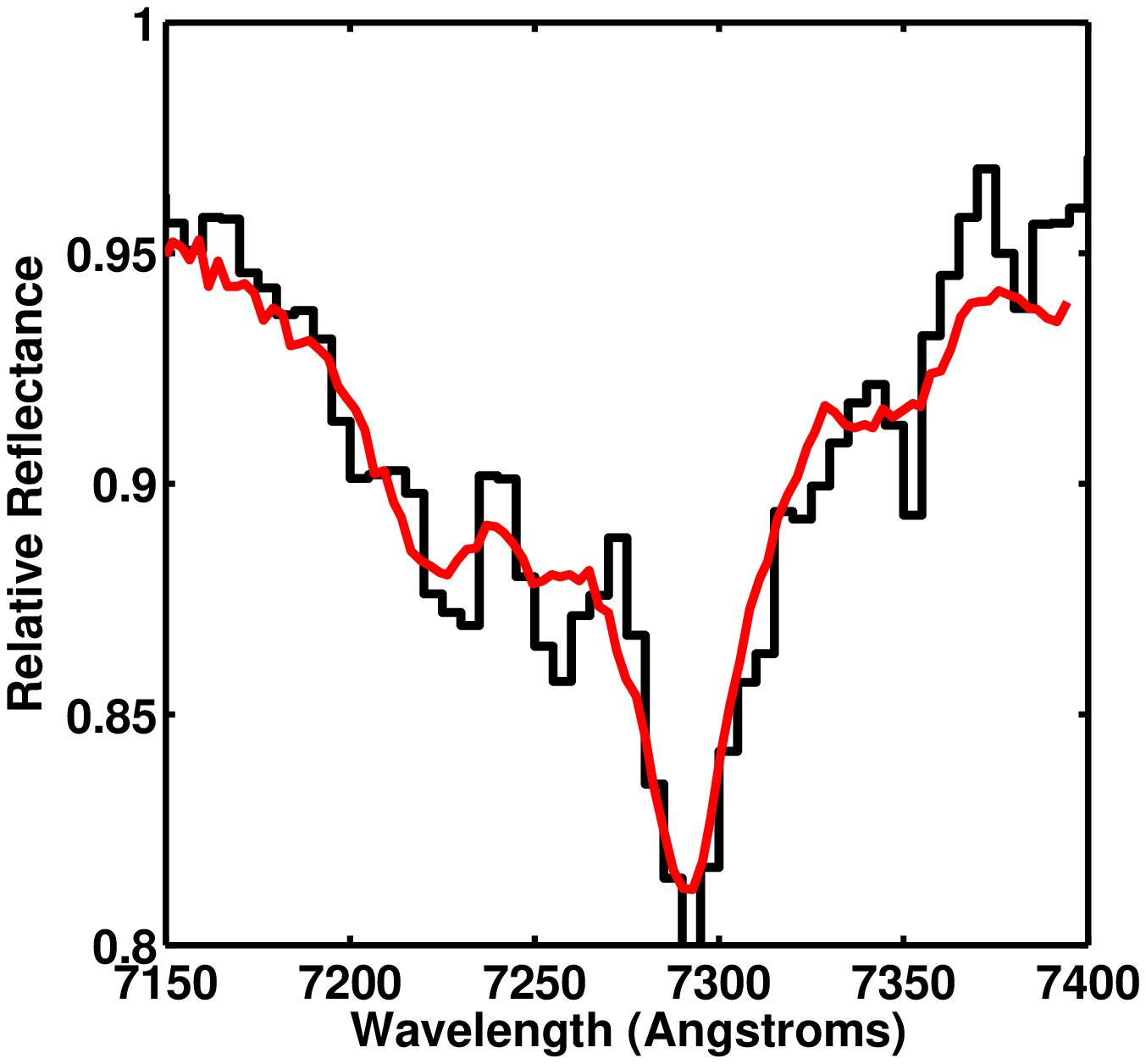}
\caption{Eris' 7296 \AA\  methane band taken on 2008 October 3 UT with the MMT 6.5 m telescope (black line), and the best fit pure methane Hapke model (red line). Blue shifting the pure methane band by 4.8 $\pm$ 1.0 \AA\  gave the best fit to Eris' band. $\chi^2$ $=$ 5}
\end{figure}

\begin{figure}
\epsscale{.70}
\plotone{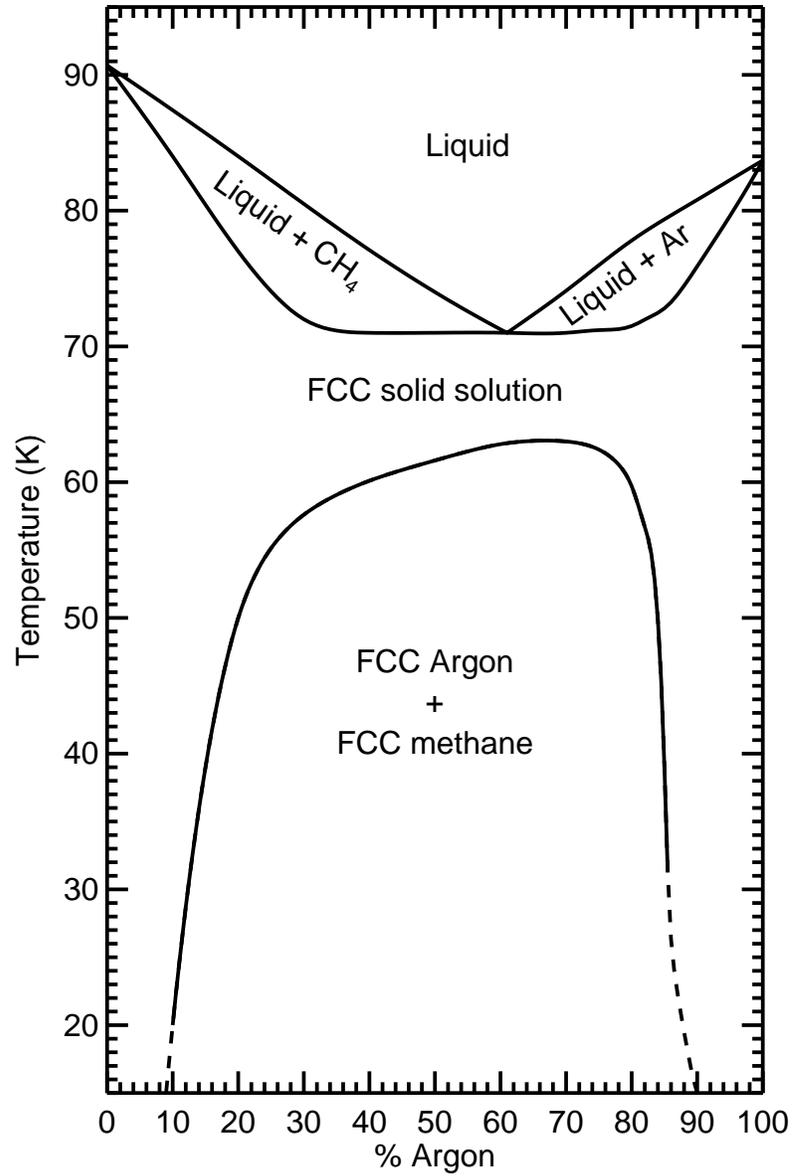}
\caption{Phase diagram for methane-argon ice mixtures \citep{gmb69}. In our experiments, we cooled an ice sample with $\sim$ 91\% argon and $\sim$ 9\% methane from 60 $^{\circ}$K to 40 $^{\circ}$K. We believe the sample went from a FCC solid solution with methane and argon mixed throughout the sample to a FCC argon $+$ FCC methane structure composed of highly methane rich regions and highly argon rich regions.}
\end{figure}

\begin{figure}
\epsscale{.70}
\plotone{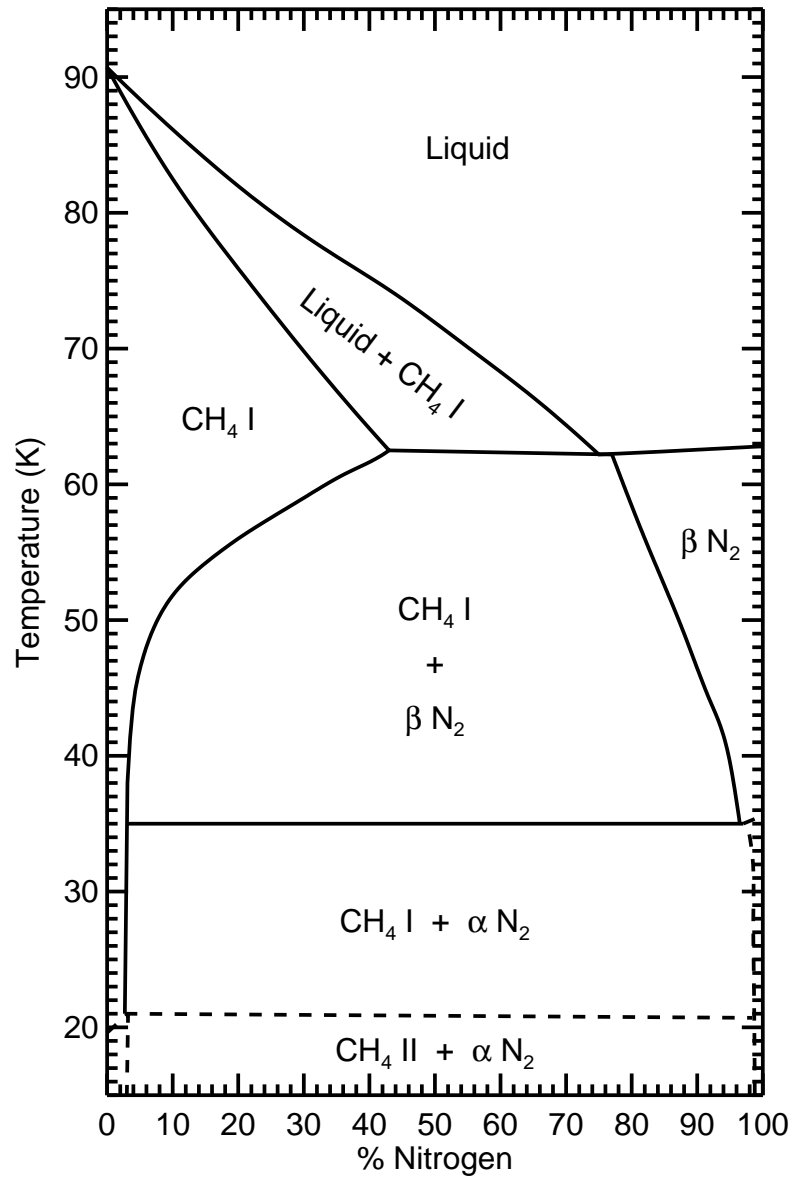}
\caption{Phase diagram for methane-nitrogen ice mixtures \citep{py83}. At the temperatures of our experiments (T $\sim$ 60 $^{\circ}$K), the samples were a solid solution of methane and nitrogen molecules (CH$_4$ I and $\beta$ N$_2$). At the temperature of Pluto (T $\sim$ 40 $^{\circ}$K), the ice probably  is composed of methane rich regions and nitrogen rich regions (CH$_4$ I $+$ $\beta$ N$_2$). At the temperature of Eris (T $\sim$ 30 $^{\circ}$K), the ice also probably  is composed of methane rich regions and nitrogen rich regions (CH$_4$ I $+$ $\alpha$ N$_2$). } 
\end{figure}

\begin{figure}
\epsscale{.80}
\plotone{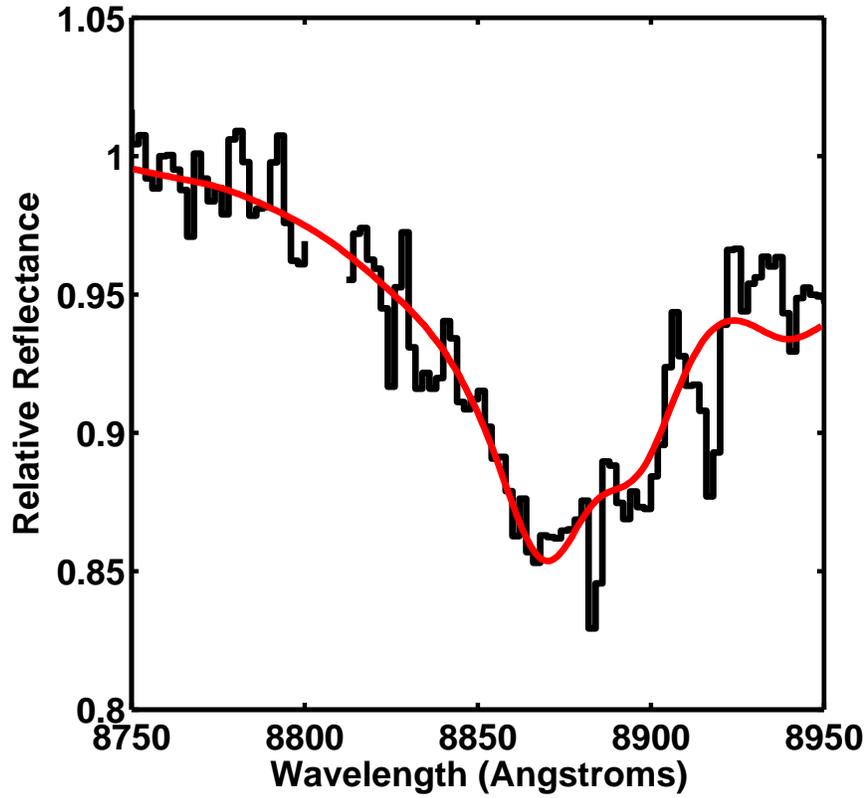}
\caption{Pluto's 8897 \AA\   methane band taken on 2007 June 21 UT with the Steward Observatory 2.3  m telescope (black line), and the best fit binary mixture Hapke model (red line). The best fit model had a methane abundance of $\sim$ 3$\%$ and a nitrogen abundance of $\sim$ 97$\%$. $\chi^2$ $=$10.}
\end{figure}

\begin{figure}
\epsscale{.80}
\plotone{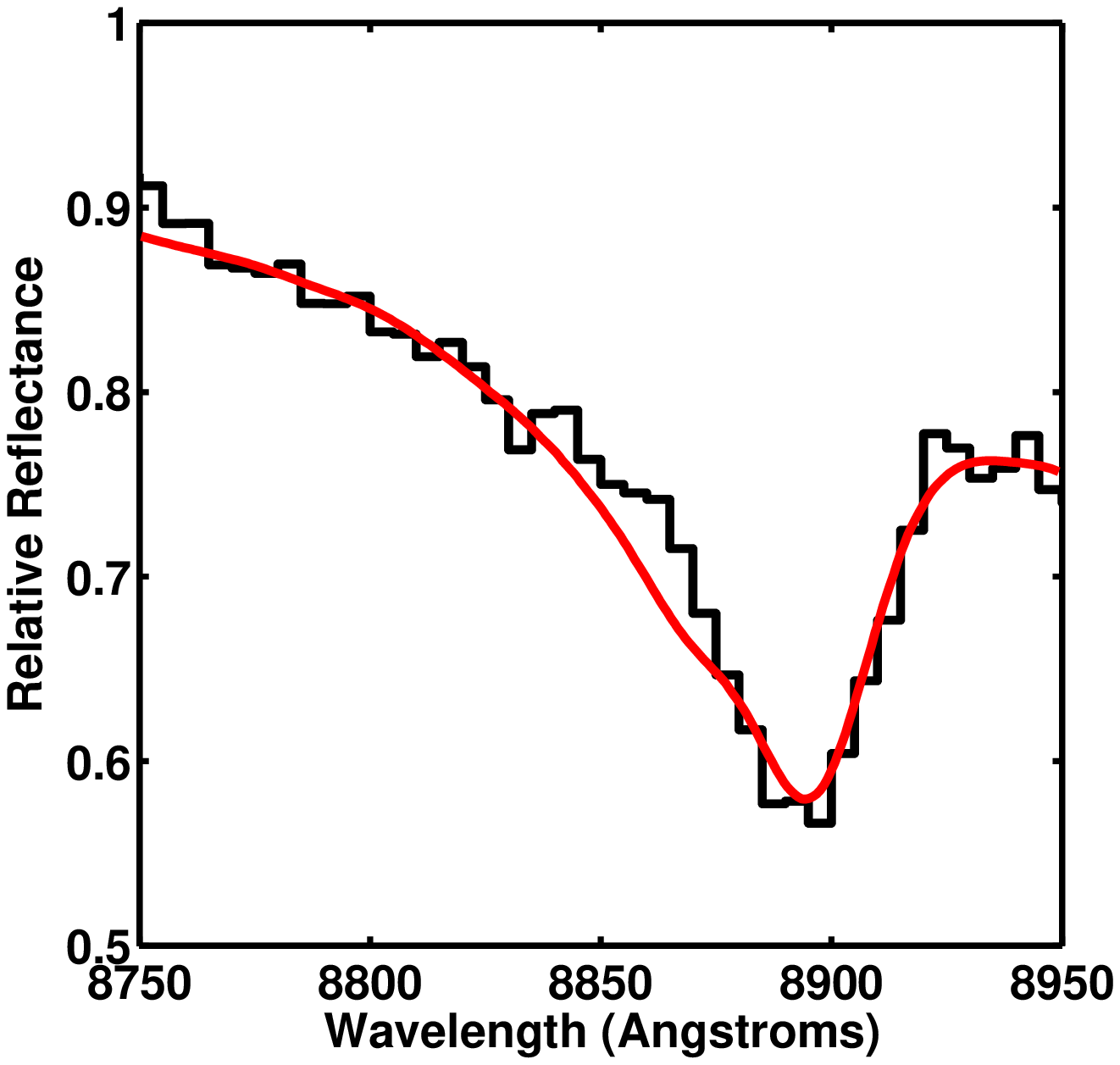}
\caption{Eris' 8897 \AA\  methane band taken on 2008 October 3 UT with the MMT 6.5 m telescope (black line), and the best fit binary mixture Hapke model (red line).  The best fit model had a methane abundance of 10$\%$ and a nitrogen abundance of 90$\%$. $\chi^2$ $=$ 10.}
\end{figure}

\begin{figure}
\epsscale{.80}
\plotone{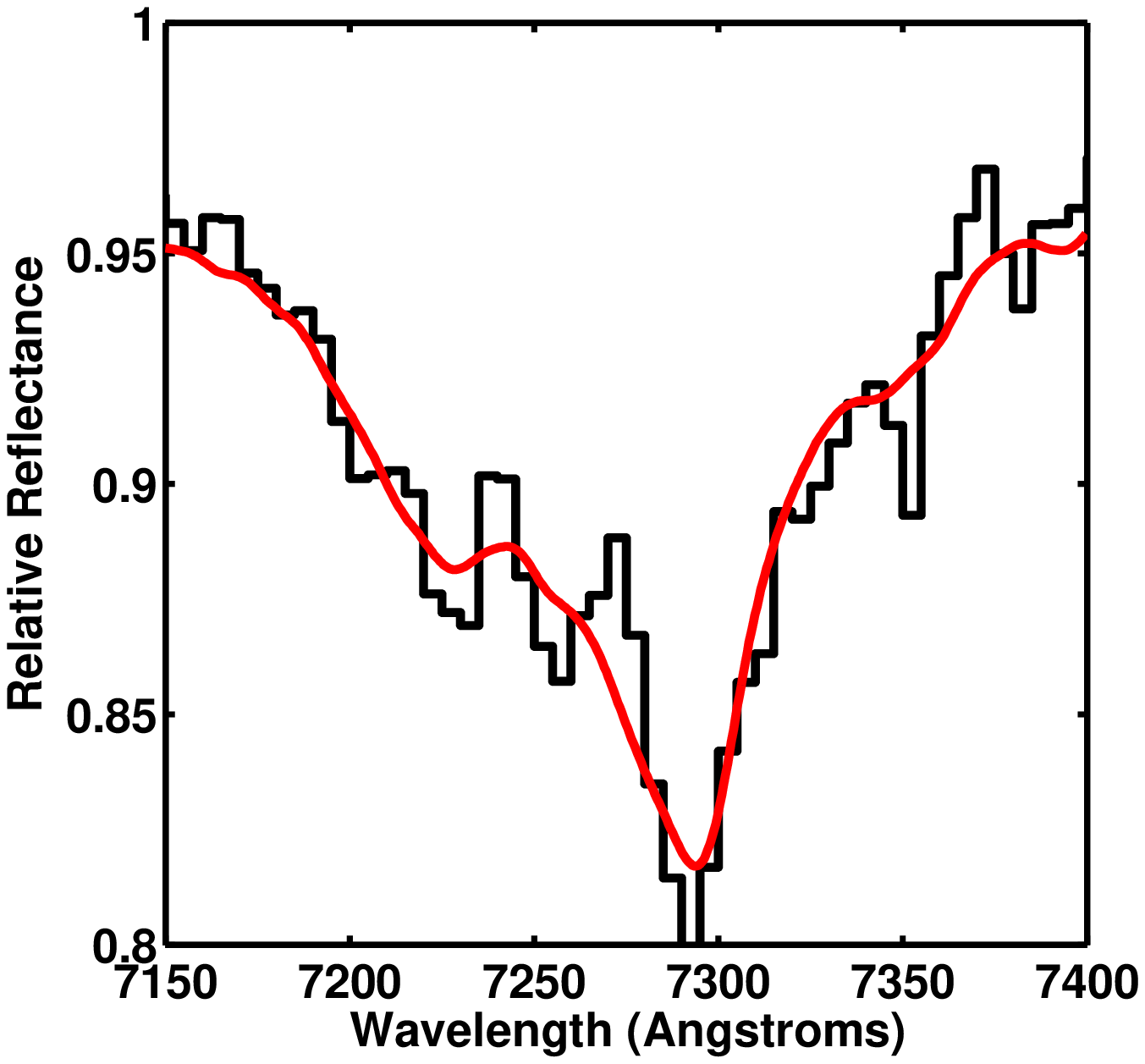}
\caption{Eris' 7296 \AA\  methane band taken on 2008 October 3 UT with the MMT 6.5 m telescope (black line), and the best fit binary mixture Hapke model (red line).  The best fit model had a methane abundance of 8$\%$ and a nitrogen abundance of 92$\%$. $\chi^2$ $=$ 3.}
\end{figure}

\begin{deluxetable}{ccc}
\tablewidth{0pt}
\tablecaption{Band Shifts For Methane-Nitrogen Samples$^a$}
\tablehead{
\colhead{ Methane$^b$} & 
\colhead{13706$^c$} & 
\colhead{11240$^c$}}

\startdata
2 & 34 &  29\\
6   & 31 & 26\\
24   & 25 & 19\\
99   & 1 & 0 \\
\enddata
\tablenotetext{a}{Samples at T $=$ 60$^{\circ}$K}
\tablenotetext{b}{Percent Methane}
\tablenotetext{c}{Shifts in units of cm$^{-1}$}
\end{deluxetable}

\clearpage

\begin{deluxetable}{ccc}
\tablewidth{0pt}
\tablecaption{Band Shifts For Methane-Argon Samples$^a$}
\tablehead{
\colhead{ Methane$^b$} & 
\colhead{13706$^c$} & 
\colhead{11240$^c$}}

\startdata
9 & 56 &  38\\
15   & 46 & 33\\
41   & 27 & 20\\
96   & 1 & 1 \\
\enddata
\tablenotetext{a}{Samples at T $=$ 60$^{\circ}$K}
\tablenotetext{b}{Percent Methane}
\tablenotetext{c}{Shifts in units of cm$^{-1}$}
\end{deluxetable}

\clearpage

\begin{deluxetable}{lccc}
\tablewidth{0pt}
\tablecaption{Band Shifts Relative to Pure Methane}
\tablehead{
\colhead{ Object} & 
\colhead{7296$^a$} & 
\colhead{8691$^a$} &
\colhead{8897$^a$}}

\startdata
Pluto & & &17 $\pm$ 2 \\
Eris & 4.8 $\pm$ 1.0 & 4.8 $\pm$ 1.1 & 4.0 $\pm$  0.6\\
\enddata
\tablenotetext{a}{Shifts in units of \AA}
\end{deluxetable}

\clearpage

\begin{deluxetable}{lccc}
\tablewidth{0pt}
\tablecaption{Methane Abundances$^a$}
\tablehead{
\colhead{ Object} & 
\colhead{7296} & 
\colhead{8691} &
\colhead{8897}}

\startdata
Pluto & & & 3.2 $\pm$ 0.2\\
Eris   & 8 $\pm$ 2 & 9 $\pm$ 2 &10 $\pm$ 2\\
\enddata
\tablenotetext{a}{Percent Methane}
\end{deluxetable}

\end{document}